\documentclass[copyright,creativecommons]{eptcs}
\providecommand{\event}{QAPL 2019}

\usepackage{graphicx}
\usepackage{url}

\providecommand{\urlalt}[2]{\href{#1}{#2}} \providecommand{\doi}[1]{doi:\urlalt{http://dx.doi.org/#1}{#1}}

\begin{document}

\title{Towards Digital Twins for the Description of Automotive Software Systems}

\author{Jan Olaf Blech
\institute{Aalto University, Finland }
\email{\quad jan.blech@aalto.fi}
}
\def\titlerunning{Digital Twins of Automotive Systems}
\def\authorrunning{J.O. Blech}

\maketitle

\begin{abstract}
We present models for automotive software that capture quantitative and qualitative aspects of software systems and the underlying hardware architecture. In particular, we consider different levels of computing power. These range from controllers up to the cloud. We present a modeling approach for software deployment taking different automotive requirements such as criticality, latency, memory, computational resources, and communication into account. Our models capture automotive software and hardware system configurations and can serve as digital twins that are digital counterparts of (usually) physical entities. Furthermore, we highlight  connected research areas and challenges. 

\end{abstract}

\section{Introduction}
In the past decades, software in the automotive domain has gained an increasingly important role: functionality that used to be realized by electronic, electrical and mechanical devices alone is now frequently controlled by software. Software can run on more than 70 ECUs (Electronic Control Units)\cite{broy2006} inside a car. In recent times, a consolidation of software and ECUs, i.e., the replacement of several microcontrollers by using a more powerful ECU, has gained increased attention. In addition new software-based functionality and the introduction of additional computational resources such as cloud-computing,  have found their way into the automotive world.  This trend is complemented by the increased use of technology that has its origins in the IT-world such as Ethernet, general purpose computing devices and operating systems such as Linux.

\begin{figure}
    \centering
    \includegraphics[width=.95\textwidth]{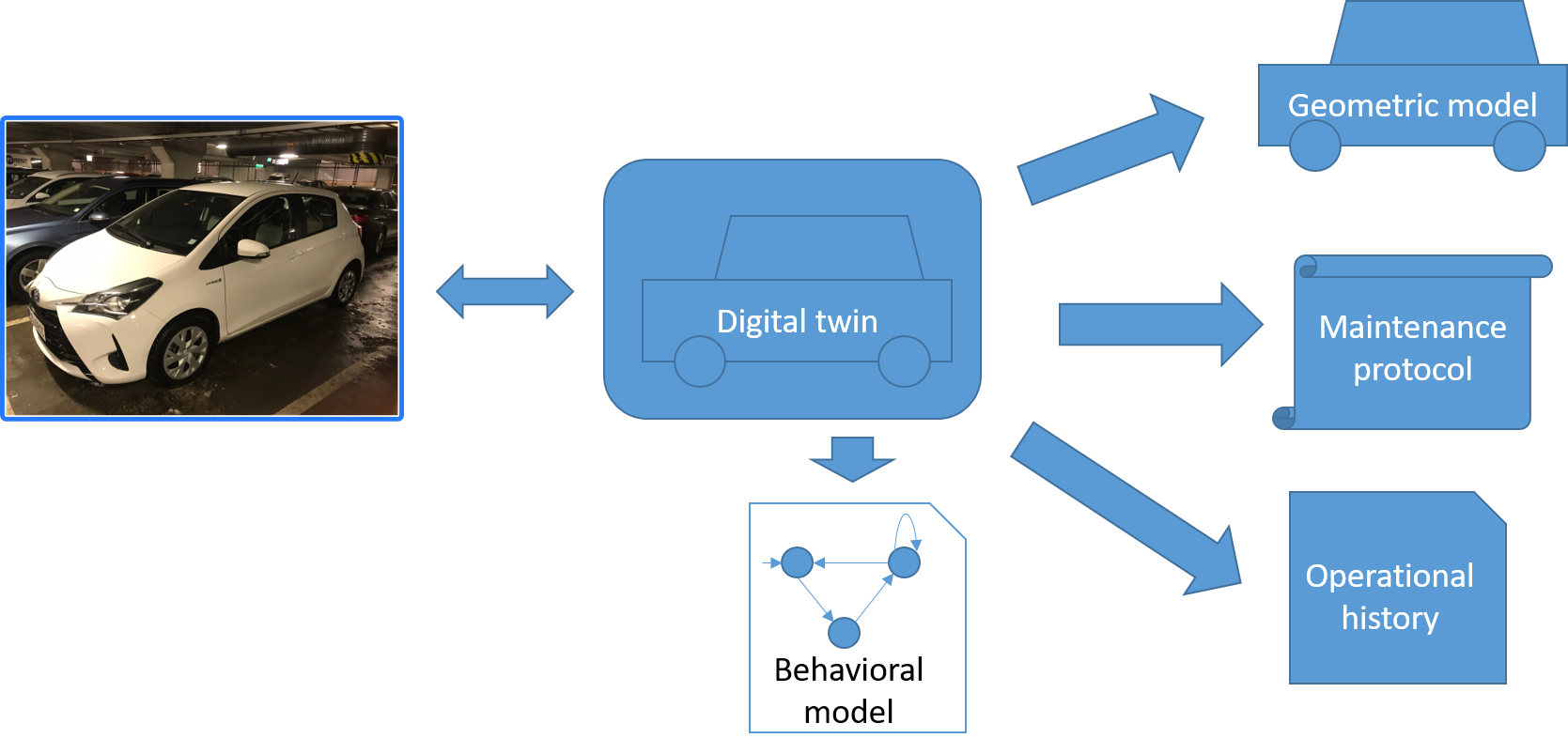}
    \caption{Digital Twin}
    \label{fig:dt}
\end{figure}

This paper primarily motivates research challenges on models serving as digital twins with a special emphasis on quantitative aspects used in the partitioning of software across different ECUs and embedded computers in a car as well as cloud-based services. The term digital twin describes a digital counterpart of a (usually) physical entity such as a product (e.g., a car), a part of a product (e.g., a car's engine) or a machine (see Figure~\ref{fig:dt}). Note, that digital twins can describe non-physical entities as well such as the software architecture of a system.

Here, we are using a notion of models that goes beyond traditional models in model-based development. Models are abstractions of systems and can include aspects such as physical characteristics, geometrical layout, wiring, communication links, hardware resources (e.g., computational power) and information on the structure of software such as components, layers and interfaces.
Models of systems such as cars can be used at design and development time, during the car's life time and even during and after the decommissioning phase. They can be used for design, documentation, optimization purposes and present an abstraction of the system and thus serve as digital twins of automotive hardware and software. 

In this paper, we use the following ingredients for interconnected computational devices:
    {\it Microcontroller}, as devices that are primarily developed to interact with sensors and actuators.
    {\it Embedded computers}, as devices that do typically feature classical IT components such as general purpose microprocessors, RAM/microprocessor on separate chips and ethernet.
    The user of a {\it cloud} service is in general unaware of the actual location and resources where the service is provided. In principle services can be shifted freely between data-centres on different continents. This full flexibility, however, may cause unpredictable delay times. The strategy to move services closer to the actual controller is sometimes referred to as fog computing \cite{fogcomputing}. If the services are located at the outer edge of the network the term edge-computing is used~\cite{edgecomputing}.

\section{Related work and Research Areas}

\begin{figure}
\centering
\includegraphics[width=5cm,angle=270]{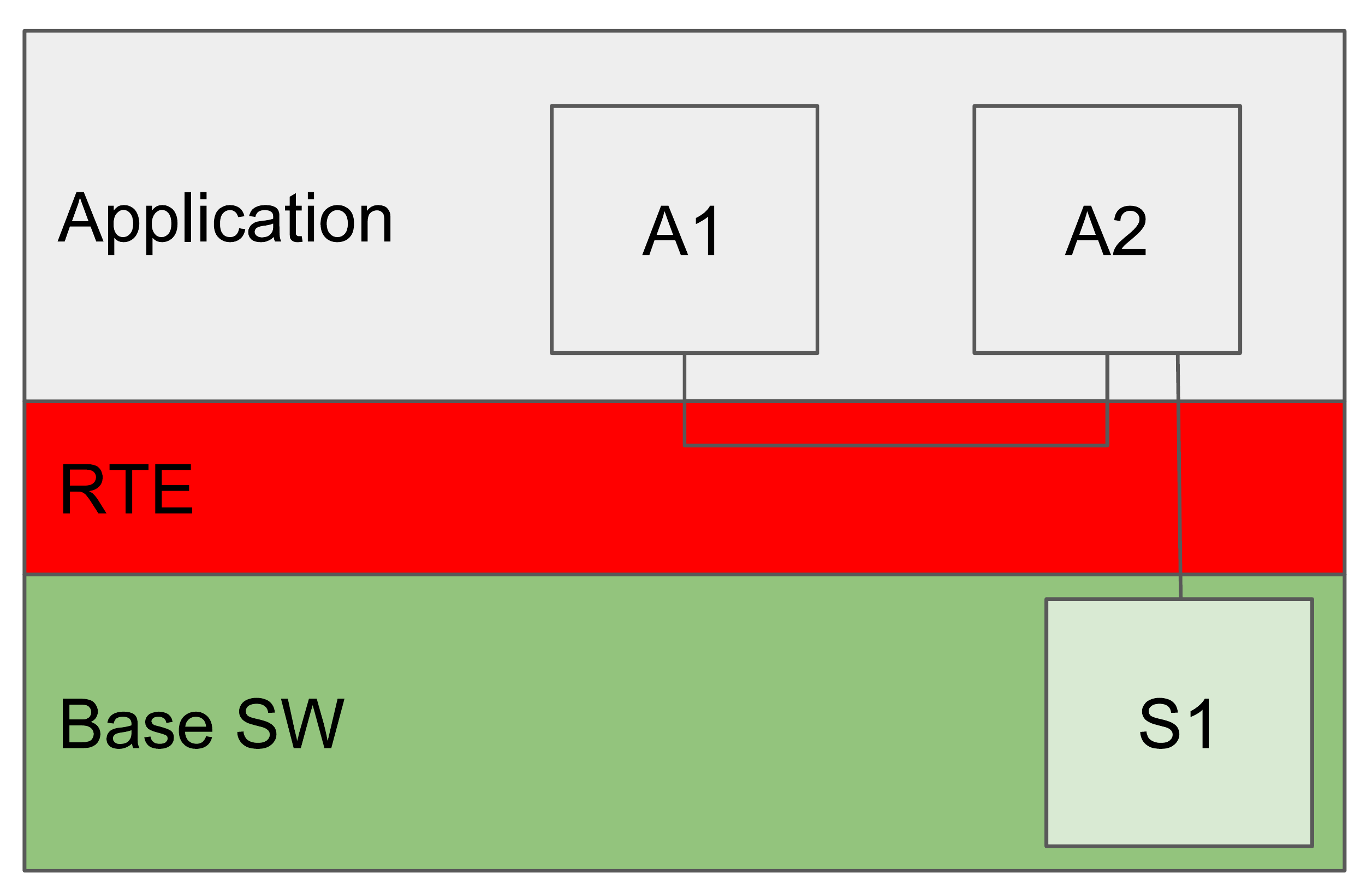}
\caption{AUTOSAR Classic Software Stack}
\label{fig:as}
\end{figure}

Typically software running on an ECU is structured in layers such as base software including drivers, and a runtime environment (RTE) which enables the deployment of application software.  Currently, the AUTOSAR Classic standard\footnote{\url{https://www.autosar.org/standards/classic-platform/}, retrieved 30th November 2018} 
and Posix-based operating systems are mainly used. In the rather monolithic AUTOSAR Classic standard, the application software itself is structured into components which may further be structured into subcomponents that comprise executable code units (called runnables in AUTOSAR). Figure~\ref{fig:as} shows the AUTOSAR Classic stack. It can be seen that all communication between different applications and services from the base software such as network communication is performed via the RTE. All communication pathes through the RTE are statically generated and can not adapt after compile time.

\begin{figure}[t]
\centering
\includegraphics[width=5.5cm,angle=270]{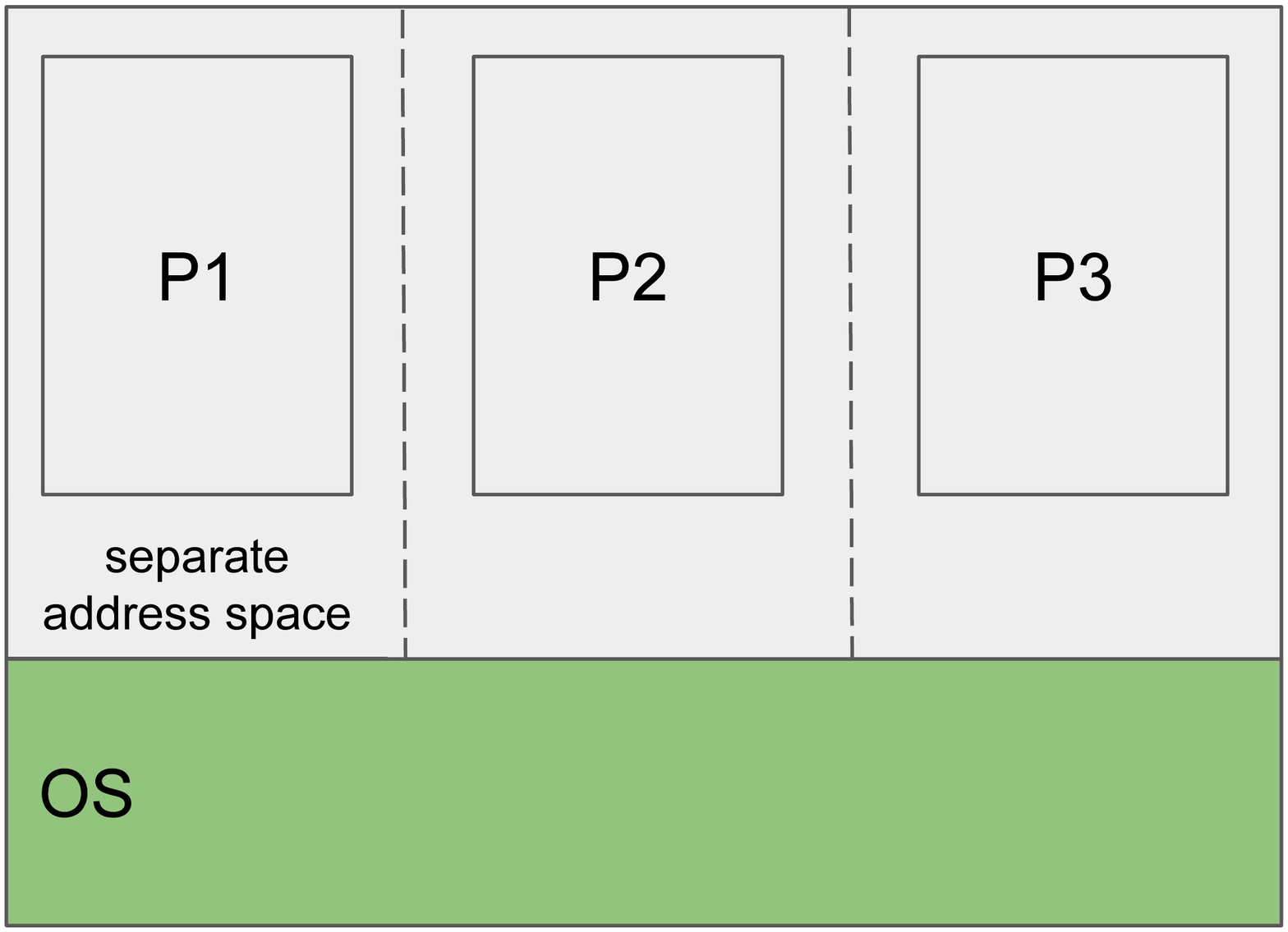}
\caption{Adaptive AUTOSAR (Posix) Software Stack}
\label{fig:as2}
\end{figure}

Posix-like operating systems such as the adaptive AUTOSAR framework (see, e.g., \cite{adaptiveautosar}) allow a greater amount of reconfiguration, typically require more resources and may have trouble to meet certain other criteria such as reliability and latency.
Figure~\ref{fig:as2} presents the architectural view of an adaptive AUTOSAR systems which is realized on-top of a Posix operating system such as Linux. In Posix systems processes (realizing applications) can be started, stopped, loaded and terminated at run-time of a system.

Mixed Criticality systems is a research area connected to our work (see e.g., \cite{mixedc}) and partitioning of software has been addressed in~\cite{ernst16}.
Scheduling is frequently regarded as an important aspect of Mixed Criticality Systems (see, e.g. 
including multi-core challenges \cite{multicoremixed}). Task scheduling in this context has been, e.g., studied in \cite{huang2011}.
Design Space Exploration in the ISO 26262 context has been studied in~\cite{schaetz15} and regards the partitioning of software components to hardware elements. The paper focuses on criticality aspects of software functions (i.e., ASIL level classification).

Hardware-Software Co-Design (e.g., \cite{hwswcodesign1} and \cite{hwswcodesign2}) regards the question whether a functionality should be realized in hardware of software, or if there is a possibility to further partition it into hardware and software parts. Design-space exploration techniques have been applied to this domain as well. 

The need for clarification of terminology in the automotive software world was proposed in \cite{broy2010} together with a service hierarchy proposition for embedded automotive software.
Behavioral types~\cite{beht1,beht2} are a related formalism that we have introduced  to describe state-based properties of both software and cyber-physical systems. The work described in this paper continues this view by presenting models for hardware and software systems. State-based behavior can be annotated using the properties featured in our models.
We have investigated the use of models serving as digital twins together with a cloud-based service infrastructure in the industrial automation domain~\cite{colleng2}. An emphasis here was on the remote response to incidents in industrial facilities.

\section{Models Mirroring the Car}
Digital twins as digital mirrors of (mostly physical) entities have gained popularity in the manufacturing domain~\cite{dtwins} especially in the Industrie 4.0~\cite{industrie40,industrie40b} context.
Models can serve as digital twins throughout the development, construction and commissioning, life-time, and decommissioning of a technical system. Meta-models are the datatypes for models. They outline what information can be kept inside a particular model.
Models can be stored in the cloud and a variety of mechanisms have been developed to maintain and use models in various stages and tools, such as the open source Eclipse Modeling Framework~\footnote{\url{https://www.eclipse.org/modeling/emf/}} that comes with modeling, tool-support, parsers and cloud-based storage mechanisms.

While we put an emphasis on software partitioning at development time, the models for digital twins introduced in this paper (both software and hardware models) can also serve the following purposes:
\begin{itemize}
    \item Reconfiguration (see e.g., \cite{wang2017} for work on model support in this context) of software at runtime (during the operation of a car). This means, that we answer the question which application software component should run on which ECU.
    \item Diagnosis, detecting malfunctions and identifying which software components are affected.
    \item Resolving issues at run-time. For example, if a software component cannot run on a particular ECU anymore (e.g., due to a hardware failure), its functionality could be shifted to another ECU or another software component could provide backup functionality.
    \item Verification and validation of constraints such as requirements on software architecture.
    \item Tracking of changes that occur during the live-time of a system. Systems may evolve during the lifetime. Maintenance protocols including the replacement of ECUs and the update history of software can be archived using models.
    \item Tracking and helping to gain relevant information during the decommissioning of a system: the history of a system may provide useful hints on how to best decommission a system.
\end{itemize}

\begin{figure} [t]
\centering
\includegraphics[width=9cm,angle=0]{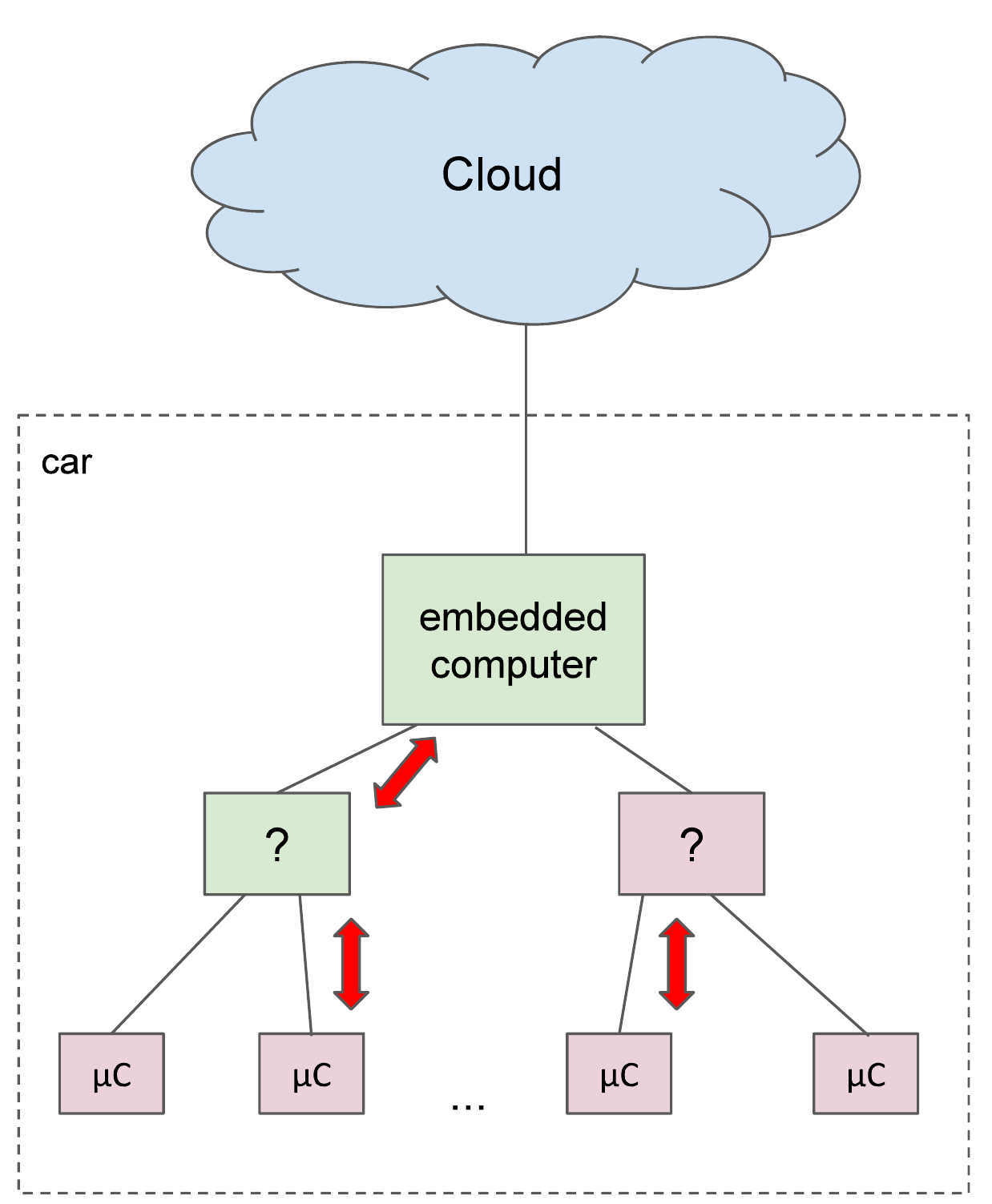}
\caption{Re-Partitioning of Software}
\label{fig:ecuhc}
\end{figure}

Figure~\ref{fig:ecuhc} shows the organization of ECUs in a car using three hierarchical levels plus the cloud. 
Shifting functionality between these levels is one goal that we want to draw attention to.

\section{Partitioning of Software}
In this paper, we are particularly interested in the question where to deploy a piece of software that realizes a functionality. In order to find an optimal location, we need to develop models that capture both the system including ECUs and physical buses as well as logical connections between different software components. Here, we present a mathematical founded model for both hardware and software.

\subsection{Proposed Hardware Model}
All devices on which software can run ($\mathit{ECU}$s) and their interconnections are modeled as a graph $(\mathit{ECU},\mathit{NL},\mathcal{P_{HW}})$ comprising a set of hardware devices $\mathit{ECU}$, a set of communication links $\mathit{NL} : \mathit{ECU} \times 2^{A_{HE}} \times \mathit{ECU}$ and a function $\mathcal{P_{HW}}: ECU \mapsto 2^{A_{ECU}}$ ($2^{A_{ECU}}$ denotes the powerset of $A_{ECU}$).
Note, that the set $\mathit{ECU}$ is quite general in order to capture all kinds of computational devices: Cloud-based services are in $\mathit{ECU}$ as well as controllers, embedded computers and  smart sensors.

\begin{figure}
    \centering
    \includegraphics[width=.85\textwidth]{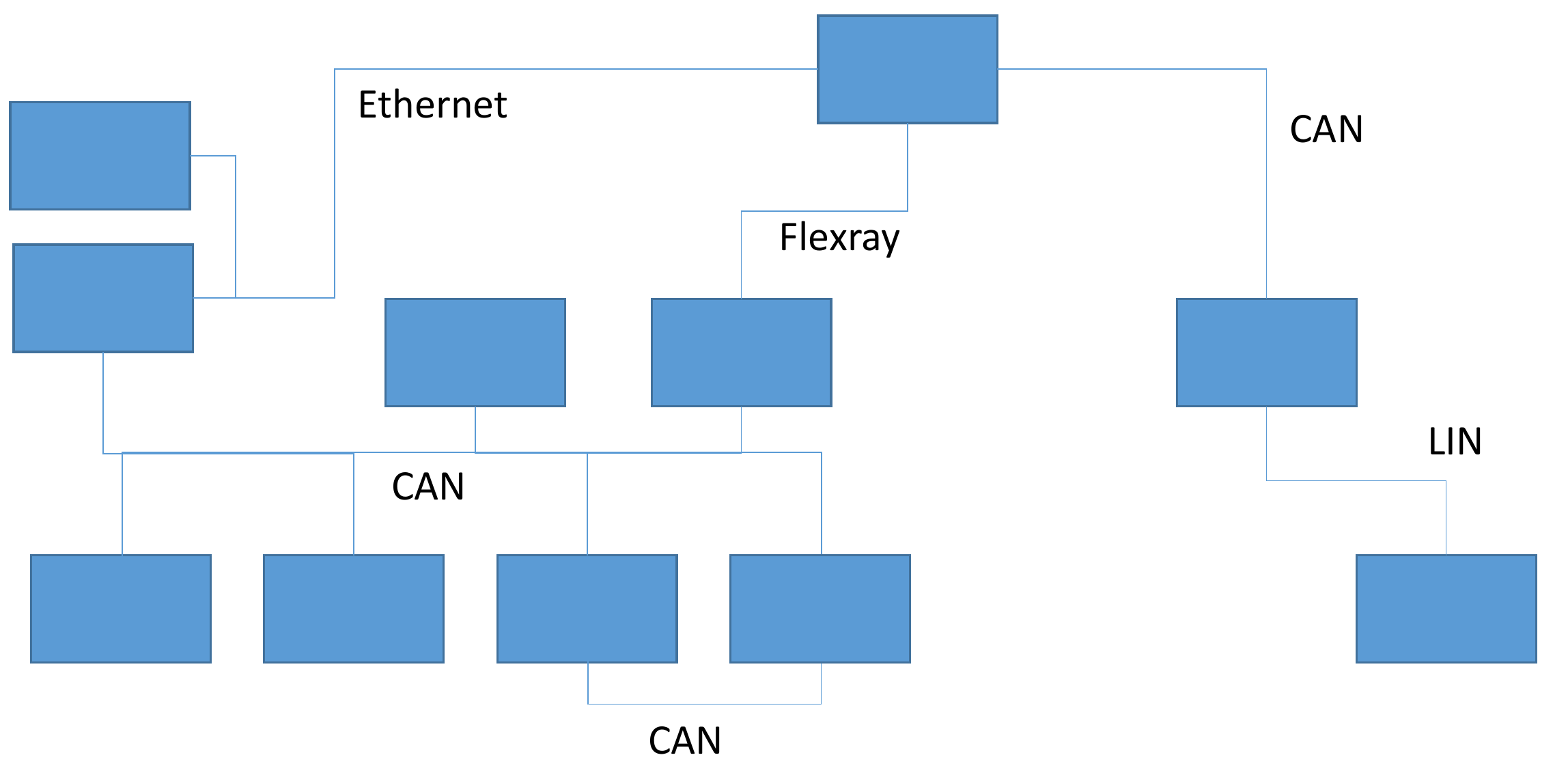}
    \caption{A Simplified View on Automotive Buses}
    \label{fig:autob}
\end{figure}

Communication links formalize hardware-based links such as buses, but also wireless communication channels between members of the $\mathit{ECU}$ set. A communication link $(ecu_i,a,ecu_j) \in \mathit{NL}$ comprises a source communication device $ecu_i$ a set of attributes $a \in 2^{A_{HE}}$ that comprises attributes such as speed and capacity and a target device $ecu_j$. Figure~\ref{fig:autob} shows a simple example for an automotive bus structure. ECUs are connected via Ethernet, Flexray, CAN and LIN buses. The figure illustrates that there can be more than one connection between two ECUs. Note, that the actual communication messages that can be sent between different ECUs are typically fixed as at an early development stage. Therefore, during the development we effectively have a one-to-one communication between different ECUs which is captured in our graph formalization.

For reasoning about a system, we need to describe properties of both computational devices and communication links. These properties are described as attributes. The link attributes are directly contained in the link, for $\mathit{ECU}s$ the function $\mathcal{P_{HW}}$ maps hardware devices to a set of attributes from $A_{ECU}$.
Typical attributes from $A_{ECU}$ comprise:
\begin{itemize}
 \item 
Computational power, such as the number of available cores, their speed, their architecture. 
\item
Hardware-Architectural features such as lockstep computation, special co-processors. 
\item
Memory characteristics such as RAM/ROM/Flash, on-chip, separate chips, size and speed. 
\item
IO capabilities such as hardware interfaces, available sensors and actuators, general purpose IO pins. 
\item
Networking capabilities, such as interfaces to specific bus and other communication infrastructure
\end{itemize}

\subsection{Proposed Software Model}
The key idea for describing a software system is to describe the application components and necessary communication between them.
The software system is described as a graph $(SWC,E,\mathcal{P_{SW}})$ featuring a set of Nodes $\mathit{SWC}$, a set of edges $E : \mathit{SWC} \times 2^{A_{SE}} \times \mathit{SWC}$ and a function $\mathcal{P_{SW}}: \mathit{SWC} \mapsto 2^{A_{SWC}}$ that maps nodes to attributes. Our model features the following abstractions:
\begin{itemize}
    \item  Software is broken down into atomic components and each atomic software component becomes a node $n_i \in SWC$.
    \item Nodes are associated with attributes to indicate resource and other hardware requirements such as RAM consumption, communication requirements resulting in links and upper bounds on communication time to actuators and sensors as well as the criticality of the functionality realized by the software component. This is realized using the function $\mathcal{P_{SW}}$ which maps a node to a set of relevant attributes $2^{A_{SWC}}$
    \item Edges $(n_j,a,n_k) \in E$ represent the required communication between atomic software components. Each edge is associated with a set of attributes $a \in 2^{A_{SE}}$. Typical attributes are the amount and nature of communication and latency requirements.
\end{itemize}
Note, that we only look at application components in our model and assume that typical requirements on the base software and run-time environments can be realized and thus exclude them from our model. In order to achieve full digital twins, additional attributes can be added to track, e.g., physical locations and maintenance protocols.

\subsection{An Example System}
To give an idea of how our models can be used, we provide a small example system.
The hardware model is given:
by a set of computational devices: $ECU = \{Cloud, GW, C1, C2 \}$, comprising the cloud $Cloud$, a gateway device realized as an embedded PC $GW$ and two controllers $C1$, $C2$.

We only regard three properties in this example: RAM, the local availability of a sensor $Sen$ and an actuator $Act$. Thus, the property function $\mathcal{P_{HW}}$ is defined as: \\ \\
    \begin{tabular} {l}
    $\mathcal{P_{HW}}(Cloud) = \{ '' RAM == infinity '' \}$ \\
    $\mathcal{P_{HW}}(GW) = \{ '' RAM == 1024 MB '' \}$ \\
    $\mathcal{P_{HW}}(C1) = \{ '' RAM == 4 MB '', Sen \}$ \\
    $\mathcal{P_{HW}}(C2) = \{ '' RAM == 2 MB '', Act \}$
    \end{tabular} \\ \\
    This means that the cloud has inifinity RAM ressources, the gateway has 1024 MB, the controller $C1$ has a local sensor $Sen$ available and 4 MB RAM and the controller $C2$ has an actuator $Act$ and 2 MB RAM.
    %
    The set of communication links is defined as: \\
    \indent $\{ (Cloud, \{\}, GW),(GW, \{\}, C1),(GW, \{\}, C2),$ \\ 
    \indent $(GW, \{\}, Cloud),(C1, \{\}, GW),(C2, \{\}, GW)\}$. \\ 
    Links between the cloud and the gateway as well as between the gateway and the controllers are contained. No properties are given in this simple example. Note, that we need to include both directions for bidiractional communication.  

\noindent The software model is defined as follows: 
\begin{enumerate}
   \item 
The set of nodes, representing software components: \\ $SWC = \{CtrlS,CtrlA,Comp1,Comp2,Comp3\}$. \\ We have five software components: $CtrlS$ is the control component for a sensor, $CtrlA$ a control component for an actuator, $Comp1$, $Comp2$ and $Comp3$ are other components performing computations or offering services. 
   \item 
 Required communication is specified as the set: \\ $\{ (CtrlS,\{\},Comp1),$  $(Comp1,\{\},CtrlA),$ \\ $(Comp1,\{\},Comp2), $ $ (Comp1,\{\},Comp3)\}$. \\
    One can see that the $CtrlS$ component needs to communicate with $Comp1$. $Comp1$ needs to communicate with $CtrlA$, $Comp2$ and $Comp3$. 
    \item 
 The properties of the software components are defined by instantiating the property mapping function $\mathcal{P_{SW}}$: \\
\begin{math}
    \begin{array}{ll}
             \mathcal{P_{SW}}(CtrlS) = \{Sen\}  &  \mathcal{P_{SW}}(Comp2) = \{\} \\
             \mathcal{P_{SW}}(CtrlA) = \{Act\}  &  \mathcal{P_{SW}}(Comp3) = \{\} \\
             \mathcal{P_{SW}}(Comp1) = \{\}  & \\
    \end{array} 
\end{math}\\
    The function is used to state requirements of the software components. Here, we have only formalized that $CtrlS$ requries the local availability of the sensor $Sen$ and $CtrlA$ requires the local availability of the actuator $Act$.
\end{enumerate}

\subsection{Goals}
The introduced models can be used to achieve at least two different goals when reasoning about good software partitoning:
\begin{enumerate}
    \item Finding the right partitioning for a given software system and a given hardware system.
    \item Finding a good hardware model for a given software system.
\end{enumerate}
To solve these tasks, we define a solution quality evaluation function: \\
$\mathcal{E}: $ \\
\indent $(\mathit{ECU},\mathit{NL},\mathcal{P_{HW}}) \times $  $(\mathit{SWC},E,\mathcal{P_{SW}}) \times $ \\
\indent $(\mathit{SWC} \mapsto ECU) $  $\mapsto int$ \\
which takes a hardware model and a software model, furthermore it takes a function $SWC \mapsto ECU$ mapping an atomic software component to an ECU. The function returns a numerical value indicating the quality of the solution. Higher numbers indicate better solutions.  Negative numbers indicate that no solution has been found.

\noindent In our example, the mapping $M : (SWC \mapsto ECU) $ given as: \\
\begin{math}
\begin{array}{ll}
M(CtrlS) = C1 & M(Comp2) = GW \\
M(CtrlA) = C2 & M(Comp3) = GW \\
M(Comp1) = GW & \\
\end{array} 
\end{math} \\
may achieve a satisfying score.The $CtrlS$ is located on $C1$. Thus, the required proximity to the sensor is given. Likewise $CtrlA$ is located on $C2$ and the required proximity to the actuator is also provided. All other components are located on the gateway. With the formalized requirements, this is a feasible solution. 
Another alternative is given below: \\
\begin{math}
\begin{array}{ll}
M(CtrlS) = C1 & M(Comp2) = Cloud\\
M(CtrlA) = C2 & M(Comp3) = Cloud \\
M(Comp1) = GW & \\
\end{array} 
\end{math} \\
Here, two of the $Comp$ components are located in the cloud. We would need to define more properties in order to determine if one solution is favorable over another.

Major research questions arise around good ways to find an appropriate mapping $M : (SWC \mapsto ECU) $.
Typically one would use design-space exploration techniques (see above) to find a solution. A simple solution is to use search-based algorithms, and search through possible mappings, by simple trying them out.
More intelligent techniques can use constraint-Solvers, e.g., SMT solvers, especially if one takes resources such as network capacity with numerical values into account.
The task gets even more challenging if the hardware model is not fixed, but has to be discovered as well.

\section{Conclusion}

We presented models with an emphasis on the partitioning of software in and around cars as well as connected research challenges and classification of work. Our mathematical founded models are a way to formally represent digital twins and take different quantitative and qualitative aspects of systems into account. They can be used during all phases in the life-cycle of a car. However, in this paper, we particularly proposed the use during design and development time.

\end{document}